\documentclass[conference]{IEEEtran}
\usepackage[]{times}
\usepackage{subfigure}
\usepackage{amsmath}
\usepackage{amssymb}
\usepackage{graphicx}
\usepackage{multirow}
\usepackage{algorithm}
\usepackage{mathrsfs}
\usepackage{url}
\usepackage[usenames]{color}
\newcommand{\sgn}{\text{sgn}}

\newtheorem{defn} {Definition}
\newtheorem{cor} {Corollary}
\newtheorem{te}{Theorem}
\newtheorem{lem}{Lemma}

\newtheorem{prop}{Proposition}

\newcommand{\mrm}{\mathrm}

\begin{document}
\title{Enhancing the Error Correction of Finite Alphabet Iterative Decoders via Adaptive Decimation}
\author{\IEEEauthorblockN{Shiva Kumar Planjery, Bane Vasi\'{c}}
\IEEEauthorblockA{Dept. of Electrical and Computer Eng.\\
University of Arizona\\
Tucson, AZ 85721, U.S.A.\\
Email: \{shivap,vasic\}@ece.arizona.edu}
\and
\IEEEauthorblockN{David Declercq}
\IEEEauthorblockA{ETIS\\
ENSEA/UCP/CNRS UMR 8051\\
95014 Cergy-Pontoise, France\\
Email: declercq@ensea.fr}
}
\maketitle
\begin{abstract}
Finite alphabet iterative decoders (FAIDs) for LDPC codes were recently shown to be capable of surpassing the Belief Propagation (BP) decoder in the error floor region on the Binary Symmetric channel (BSC). More recently, the technique of decimation which involves fixing the values of certain bits during decoding, was proposed for FAIDs in order to make them more amenable to analysis while maintaining their good performance. In this paper, we show how decimation can be used adaptively to further enhance the guaranteed error correction capability of FAIDs that are already good on a given code. The new adaptive decimation scheme proposed has marginally added complexity but can significantly improve the slope of the error floor performance of a particular FAID. We describe the adaptive decimation scheme particularly for 7-level FAIDs which propagate only 3-bit messages and provide numerical results for column-weight three codes. Analysis suggests that the failures of the new decoders are linked to stopping sets of the code. 

\end{abstract}
%
\section{Introduction}\label{sect_Intro}
The error floor problem of low-density parity-check (LDPC) codes under iterative decoding is now a well-known problem, where the codes suffer from an abrupt  degradation in their error-rate performance in spite of having good minimum distance. The problem has been attributed to the presence of harmful configurations generically termed as {\it trapping sets} \cite{richardsontrap} present in the Tanner graph, which cause the iterative decoder to fail for some low-noise configurations, thereby reducing its guaranteed error correction capability to an extent that is far from the limits of maximum likelihood decoding. More importantly for the BSC, the slope of the error floor is governed by the guaranteed correction capability \cite{milos}.

Recently, a new class of finite alphabet iterative decoders (FAIDs) that have a much lower complexity than the BP decoder, were proposed for LDPC codes on the BSC \cite{planjeryisit2010} \cite{FAIDjournal} and were shown to be capable of outperforming BP in the error floor. Numerical results on several column-weight three codes showed that there exist 7-level FAIDs requiring only 3 bits of precision that can achieve a better guaranteed error correction ability than BP, thereby surpassing it in the error floor region. However, analyzing these decoders for providing performance guarantees proved to be difficult. 

More recently, decimation-enhanced FAIDs \cite{planjeryisit2011} were proposed for BSC in order to make FAIDs more amenable to analysis while maintaining their good performance. The technique of decimation involves guessing the values of certain bits, and fixing them to these values while continuing to estimate the remaining bits (see \cite{planjeryisit2011} for references). In \cite{planjeryisit2011}, the decimation was carried out by the FAID based on messages passed for some iterations, and a decimation scheme was provided such that a 7-level DFAID matched the good performance of the original 7-level FAID while being analyzable at the same time.  

In this paper, we show how decimation can be used adaptively to further increase the guaranteed error correction capability of FAIDs. The adaptive scheme has only marginally increased complexity, but can significantly improve the error-rate performance compared to the FAIDs. We specifically focus on decoders that propagate only 3-bit messages and column-weight three codes since these enable simple implementations and thus have high practical value. We also provide some analysis of the decoders which suggests that the failures are linked to stopping sets of the code. Numerical results are also provided to validate the efficacy of the proposed scheme.

\section{Preliminaries}\label{sect_Prelim}
Let $G=(V\cup C,E)$ denote the Tanner graph of an ($n$,$k$) binary LDPC code $\mathcal{C}$ with the set of variable nodes $V=\{v_1,\cdots,v_n\}$ and set of check nodes $C=\{c_1,\cdots,c_m\}$. $E$ is the set of edges in $G$. A code $\mathcal{C}$ is said to be $d_v$-left-regular if all variable nodes in $V$ of graph $G$ have the same degree $d_v$. The degree of a node is the number of its neighbors. $d_{min}$ is the minimum distance of the code $\mathcal{C}$. 

A trapping set is a non-empty set of variable nodes in $G$ that are not eventually corrected by the decoder \cite{richardsontrap}.  

A multilevel FAID $\mathscr{F}$ is a 4-tuple given by $\mathscr{F}=(\mathcal{M},\mathcal{Y},\Phi_v,\Phi_c)$ \cite{FAIDjournal}. The messages are levels confined to an alphabet $\mathcal{M}$ of size $2s+1$ defined as $\mathcal{M}=\{ 0,\pm L_i \ :1\leq i\leq s  \}$, where $L_i\in\mathbb{R^{+}}$ and $L_i>L_j$ for any $i>j$. $\mathcal{Y}$ denotes the set of possible {\it channel values} that are input to the decoder. For the case of BSC as 
$\mathcal{Y}=\{\pm \mrm{C}: \mrm{C}\in\mathbb{R^{+}}\}$, and for each variable node $v_i\in V$, the channel value $y_i\in \cal{Y}$ is determined by $y_i=(-1)^{r_i}\mrm{C}$, where $r_i$ is received from the BSC at $v_i$. 

Let $m_1,\cdots,m_u$ denote the incoming messages to a node. The update function $\Phi_c: \mathcal{M}^{d_c-1} \to \mathcal{M} $ is used at a check node with degree $d_c$, and is defined as
\begin{displaymath}
\Phi_c(m_1,\ldots,m_{d_c-1}) = \left(\prod_{j=1}^{d_c-1}\sgn(m_j)\right) \min_{j \in \{1,\ldots,d_c-1\}}(|m_j|)
\end{displaymath}
where $\sgn$ denotes the standard signum function. 

The update function $\Phi_v:\mathcal{Y} \times \mathcal{M}^{d_v-1} \to \mathcal{M} $ is a symmetric rule used at a variable node with degree $d_v$ and is defined as
\begin{displaymath}
\Phi_v(y_i,m_1,m_2,\cdots,m_{d_v-1})=Q\left(\sum_{j=1}^{d_v-1}m_j+ \omega_i \cdot y_i \right).
\end{displaymath}
The function $Q$ is defined based on a threshold set $\mathscr{T}=\{T_i  :1\leq i\leq s+1 \}$ where $T_i\in\mathbb{R^{+}}$ and $T_i>T_j$ if $i>j$, and $T_{s+1}=\infty$, such that $Q(x)=\mrm{sgn}(x)L_i$  if $T_{i}\leq |x| < T_{i+1}$, and $Q(x)=0$ otherwise. The weight $\omega_i$ is computed at node $v_i$ using a symmetric function $\Omega:\mathcal{M}^{d_v-1}\to \{0,1\}$. Based on this, $\Phi_v$ can be described as a linear-threshold (LT) or non-linear threshold (NLT) function. If $\Omega=1$ (or constant), then it is an LT function, else it is an NLT function. $\Phi_v$ can also be described as a look-up table (LUT) as shown in Table \ref{LUT244325} (for $y_i=-\mrm{C}$, it can be obtained from symmetry). This rule is an NLT rule and will be one of the rules used in the proposed decimation scheme. 

\begin{table}[tp]
	\centering
  \caption{$\Phi_v$ of a 7-level FAID for $y_i=+\mrm{C}$ for a code with $d_v=3$}
	  
		\begin{tabular}{|c|c|c|c|c|c|c|c|}
		\hline $m_1\backslash m_2$ & -$L_3$ & -$L_2$ & -$L_1$ & 0  & $L_1$ & $L_2$ & $L_3$\\
		\hline 	-$L_3$	 &	-$L_3$	&	-$L_3$	&	-$L_2$	&	-$L_1$	&	-$L_1$	&	-$L_1$	&	$L_1$\\
		\hline	-$L_2$	 &	-$L_3$	&	-$L_1$	&	-$L_1$	&	0	&	$L_1$	&	$L_1$	&	$L_3$\\
		\hline	-$L_1$	&	-$L_2$	&	-$L_1$	&	0	&	0	&	$L_1$	&	$L_2$	&	$L_3$\\
		\hline	0				&	-$L_1$	&	0	&	0	&	$L_1$	&	$L_2$	&	$L_3$	&	$L_3$\\
		\hline	$L_1$		&	-$L_1$	&	$L_1$	&	$L_1$	&	$L_2$	&	$L_2$	&	$L_3$	&	$L_3$\\
		\hline	$L_2$		&	-$L_1$	&	$L_1$	&	$L_2$	&	$L_3$	&	$L_3$	&	$L_3$	&	$L_3$\\
		\hline	$L_3$		&	$L_1$	&	$L_3$	&	$L_3$	&	$L_3$	&	$L_3$	&	$L_3$	&	$L_3$\\
		\hline
		\end{tabular}
	
\label{LUT244325}
\end{table}

Let $\mathcal{T}_{i}^k(G)$ denote the computation tree of graph $G$ corresponding to a decoder $\mathscr{F}$ enumerated for $k$ iterations with variable node $v_i\in V$ as its root. A node $w\in \mathcal{T}_{i}^k(G)$ is a {\it descendant} of a node $u\in \mathcal{T}_{i}^k(G)$ if there exists a path starting from node $w$ to the root $v_i$ that traverses through node $u$. 

\begin{defn}
[Isolation assumption] Let $H$ be a subgraph of $G$ induced by $P\subseteq V$ with check node set $W\subseteq C$. The computation tree $\mathcal{T}_i^k(G)$ with the root $v_i\in P$ is said to be isolated if and only if for any node $u\notin P\cup W$ in $\mathcal{T}_i^k(G)$, $u$ does not have any descendant belonging to $P\cup W$. If $\mathcal{T}_i^k(G)$ is isolated $\forall v_i \in P$, then the subgraph $H$ is said to satisfy the isolation assumption in $G$ for $k$ iterations.
\end{defn}

{\it Remark:} The above definition is a revised version of the one given in \cite{planjeryisit2010}.

The {\it critical number} of a FAID $\mathcal{F}$ on a subgraph $H$ is the smallest number of errors for which $\mathcal{F}$ fails on $H$ under the isolation assumption.  

Let $\mathcal{N}(u)$ denote the set of neighbors of a node $u$ in the graph $G$ and let $\mathcal{N}(U)$ denote the set of neighbors of all $u\in U$. Let $m_k(v_i,\mathcal{N}(v_i))$ denote the set of outgoing messages from $v_i$ to all its neighbors in the $k^{th}$ iteration. Let $b_i^{k}$ denote the bit associated to a variable node $v_i\in V$ that is decided by the iterative decoder at the end of the $k^{th}$ iteration.

\section{Adaptive decimation-enhanced faids}\label{sect_faid}
We will first provide some definitions and notations related to the concept of decimation and discuss its incorporation into the framework of FAIDs before we delve into the details of adaptive decimation. 
\begin{defn}
A variable node $v_i$ is said to be {\it decimated} at the end of $l^{th}$ iteration if $b_i^{k}$ is set to $b_i^{\ast}$ $\forall k\geq l$. Also, 
$m_k(v_i,\mathcal{N}(v_i))=\{(-1)^{b_i^{\ast}}L_s\}$, $\forall k\geq l$ irrespective of its incoming messages, i.e., $v_i$ always sends the strongest possible messages. 
\end{defn}

{\it Remark:} If a node $v_i$ is decimated, then all its descendants in the computation tree $\mathcal{T}_i^k(G)$ can be deleted since the node always sends $(-1)^{b_i^{\ast}}L_s$ to its parent.

A \textit{decimation rule} $\beta:\mathcal{Y} \times \mathcal{M}^{d_v} \to \{-1,0,1\}$ is a function used by the decoder to decide whether a variable node should be decimated and what value it should be decimated to. Let $\gamma_i$ denote the output of a decimation rule applied to a node $v_i$. If $\gamma_i=0$, then the node is not decimated. If $\gamma_i= 1$, then $b_i^{\ast}=0$, and if $\gamma_i=-1$, then $b_i^{\ast}=1$

{\it Remark:} The decimation rule is a function of the channel value and the most recent incoming messages received by the variable node before the decimation rule is applied. 

We shall refer to each instance of applying a decimation rule on all the variable nodes as a {\it decimation round}. 

There are two key aspects to note regarding the application of a decimation rule: 1) the decimation rule is applied after messages are passed iteratively for some $l$ iterations, and 2) after each instance of applying the decimation rule, all messages are cleared to zero (which is practically restarting the decoder except that the decimated nodes remain decimated). 

Let $N_d$ denote the number of decimation rounds carried out by the decoder with a given decimation rule $\beta$ beyond which no more variable nodes are decimated. 

\begin{defn}
The {\it residual graph} $G'$ is the induced subgraph of the set of variable nodes in $G$ that are not decimated after $N_d$ decimation rounds.
\end{defn}

We can now formally define the class of adaptive decimation-enhanced multilevel FAIDs (ADFAIDs) as follows. A decoder belonging to such a class denoted by $\mathscr{F}^{A}$ is defined as  $\mathscr{F}^A=(\mathcal{M},\mathcal{Y},\Phi_v^{D},\Phi_v^{d},\Phi_v^{r},\mathcal{B},\Phi_c)$, where the sets
$\mathcal{M}$ and $\mathcal{Y}$, and the map $\Phi_c$ are same as the ones defined for a multilevel FAID. The map $\Phi_v^{D}:\mathcal{Y} \times \mathcal{M}^{d_v-1} \times \{-1,0,1\} \to \mathcal{M}$ is the update rule used at the variable node. It requires the output of a decimation rule $\beta$ as an one of its arguments and also uses the maps $\Phi_v^{d}:\mathcal{Y} \times \mathcal{M}^{d_v-1} \to \mathcal{M}$ and $\Phi_v^{r}:\mathcal{Y} \times \mathcal{M}^{d_v-1} \to \mathcal{M}$ to compute its output. For simplicity, we define it for the case of $d_v=3$ as follows.
\begin{displaymath}
\Phi_v^{D}(y_i,m_1,m_2,\gamma_i)=\Bigg\{ \begin{tabular}{ll}
 $\Phi_v^{d}(y_i,m_1,m_2)$,& if $\gamma_i=0,p\leq N_d$ \\
 $\Phi_v^{r}(y_i,m_1,m_2)$,& if $\gamma_i=0,p>N_d$\\
$\gamma_i L_s$,& if $\gamma_i=\pm 1$ \\
\end{tabular}
\end{displaymath}
where $p$ denotes the $p^{th}$ decimation round completed by the decoder. The maps $\Phi_v^{d}$ and $\Phi_v^{r}$ are defined as either LT or NLT functions or as look-up tables similar to $\Phi_v$ of a FAID $\mathscr{F}$. 

{\it Remark:} The new class of decoders proposed in this paper use two different maps, $\Phi_v^{d}$ and $\Phi_v^{r}$, for updating the messages on non-decimated variable nodes. $\Phi_v^{d}$ is the map used to update messages specifically during the decimation procedure, whereas $\Phi_v^{r}$ is the map used to decode the remaining non-decimated nodes after the decimation procedure is completed. Also note that for the case of $|\mathcal{M}|=7$, we restrict the definition of $\Phi_v^{d}$ to satisfy $\Phi_v^{d}(\mrm{C},0,0)=L_1$, $\Phi_v^{d}(\mrm{C},L_1,L_1)=L_2$, and $\Phi_v^{d}(\mrm{C},L_2,L_2)=L_3$. $\Phi_v^{r}$ is also defined similarly. 

\begin{prop}
Given a decimation rule $\beta$, if the number of decimated nodes after the $p^{th}$ decimation round is the same as the number of decimated nodes after the $(p-1)^{th}$ decimation round, then no additional nodes will get decimated in the subsequent decimation rounds. 
\end{prop}

{\it Remark:} This would be the stopping criterion used for the decimation procedure. In the above case, $N_d=p$.

The set $\mathcal{B}$ is the set of decimation rules used for adaptive decimation and for any $\beta \in \mathcal{B}$, it satisfies the following properties (specified for $d_v=3$).
\begin{enumerate}
\item $\beta(\mrm{C},m_1,m_2,m_3)=-\beta(-\mrm{C},-m_1,-m_2,-m_3)$ \\ $\forall m_1,m_2,m_3\in \mathcal{M}$
\item $\beta(\mrm{C},m_1,m_2,m_3)\neq-1$ and $\beta(-\mrm{C},m_1,m_2,m_3)\neq1$ $\forall m_1,m_2,m_3\in \mathcal{M}$
\item Given $m_1,m_2,m_3\in\mathcal{M}$, if $\beta(\mrm{C},m_1,m_2,m_3)=1$, then $\beta(\mrm{C},m_1^{\prime},m_2^{\prime},m_3^{\prime})=1$ 
$\forall m_1^{\prime},m_2^{\prime},m_3^{\prime} \in \mathcal{M}$ such that $m_1^{\prime}\geq m_1$, $m_2^{\prime}\geq m_2$, and $m_3^{\prime}\geq m_3$. 
\end{enumerate}

{\it Remark:} Property 2 implies that a node $v_i$ can be decimated to zero only if $y_i=\mrm{C}$ and to one only if $y_i=-\mrm{C}$. Consequently a node initially correct will never be decimated to a wrong value and a node initially wrong will never be decimated to the correct value.  Then, a necessary condition for successful decoding is that no node initially in error is decimated. We shall now restrict our discussion to $d_v=3$ for the remainder of the paper. 

For a given decimation rule $\beta$, a set $\Xi$ can be used to completely specify $\beta$, where $\Xi$ is defined as the set of all unordered triples $(m_1,m_2,m_3)\in \mathcal{M}^3$ such that $\beta(\mrm{C},m_1,m_2,m_3)=1$. Note that for any unordered triple $(m_1,m_2,m_3)\in \Xi$, $\beta(\mrm{-C},-m_1,-m_2,-m_3)=-1$ by property 1, so $\Xi$ is sufficient to completely specify $\beta$. A $\beta$ is considered to be a {\it conservative} decimation rule if  $|\Xi|$ is small and an {\it aggressive} rule if $|\Xi|$ is large. 

Note that the class of decimation-enhanced FAIDs defined in our previous work \cite{planjeryisit2011} is a special case of the newly proposed decoders where $\Phi_v^{d}=\Phi_v^{r}$ and $\mathcal{B}=\{\beta\}$. In other words, only a single non-adaptive decimation rule and a single map is used for updating messages in the DFAIDs of \cite{planjeryisit2011}. 
 
For the remainder of the paper, we shall refer to variable nodes that are initially in error in $G$ as {\it error} nodes and variable nodes that are initially correct as {\it correct} nodes.

\subsection{Motivation for adaptive decimation}
Given an error pattern of relatively low weight ($\leq\lfloor\frac{d_{min}-1}{2}\rfloor$), the primary role of decimation is to isolate the subgraph associated with the error pattern from the rest of the graph by decimating as many correct nodes outside this subgraph as possible.  The rationale behind resetting the messages to zero at the end of each decimation round is to allow more non-decimated correct nodes that are close to the neighborhood of the decimated correct nodes to possibly be decimated as long as none of the error nodes have been decimated. This is possible since the decimated nodes always send the strongest message ($\pm L_s$). 

Now if a given error pattern is such that the error nodes are relatively clustered with many interconnections between them through their neighboring check nodes, then a more conservative $\beta$ would have to be used by the decoder to ensure that none of the error nodes are decimated. However, if the error pattern is such that the error nodes are more spread out, then it may be desirable to use a more aggressive $\beta$ as there will be many correct nodes in the neighborhood of  the error nodes that can be decimated without decimating the error nodes, and, in turn, possibly help the decoder to converge. This is our main motivation for the use of adaptive decimation in the newly proposed decoders, and we will eventually show that adaptive decimation can help achieve an increase in the guaranteed error correction capability of the code. 

\subsection{Proposed scheme} 

We will now describe a particular adaptive decimation scheme used by the decoder $\mathscr{F}^{A}$ in order to enhance the guaranteed error correction capability. In the proposed scheme, the set $\mathcal{B}$ consists of two decimation rules, namely $\mathcal{B}=\{\beta^{(1)},\beta^{(2)}\}$, where $\Xi^{(1)}$ and $\Xi^{(2)}$ are the sets of unordered triples that completely specify the rules $\beta^{(1)}$ and $\beta^{(2)}$ respectively. The rule $\beta^{(1)}$ is used only once at the end of the third iteration, and then from that point, $\beta^{(2)}$ is used after every two iterations ($l=2$). The use of adaptive decimation is carried out only through $\beta^{(2)}$ as follows.  

We define a sequence of decimation rules $\{\beta^{(2)[j]}\}_j$ from $\beta^{(2)}$ by considering ordered subsets of $\Xi^{(2)}$ with increasing size. Let $N_{\beta}$ be the number of rules in the sequence $\{\beta^{(2)[j]}\}_j$ and let $\Xi^{(2)[j]}$ denote the set that specifies the rule $\beta^{(2)[j]}$. Then $\Xi^{(2)[j]}$ is defined for each $\beta^{(2)[j]}$ in a way such that $\Xi^{(2)[j]} \subset \Xi^{(2)[j+1]}$ $\forall i\in\{1,\ldots,N_{\beta}-1\}$ with $\Xi^{(2)[N_{\beta}]}=\Xi^{(2)}$. This implies that the sequence of rules are such that $\beta^{(2)[j+1]}$ is less conservative than $\beta^{(2)[j]}$, with $\beta^{(2)[1]}$ being the most conservative and $\beta^{(2)[N_\beta]}=\beta^{(2)}$ being least conservative (or most aggressive). Note that each subset $\Xi^{(2)[j]}$ must be chosen in a manner that ensures that its corresponding rule $\beta^{(2)[j]}$ satisfies the properties of $\beta$ mentioned previously.

For a given error pattern, the decoder starts the decimation procedure by passing messages using the map $\Phi_v^{d}$ and applying the decimation rule $\beta^{(1)}$ at the end of the third iteration after which the messages are reset to zero. Then the most conservative rule in the sequence $\{\beta^{(2)[j]}\}_j$, which is $\beta^{(2)[1]}$, is used after every two iterations (followed by resetting the messages) until no more nodes can be decimated. The map $\Phi_v^{r}$ then is used to decode the remaining non-decimated nodes. If the decoder still does not converge, then the whole decoding process is repeated by using a more aggressive rule $\beta^{(2)[2]}$ in place of $\beta^{(2)[1]}$. This decoding process continues until the decoder converges or until all rules in the sequence $\{\beta^{(2)[j]}\}_j$ have been used. Let $N_b$ denote the number of decimated bits at the end of a decimation round. The decoding scheme can be summarized as follows. Note that this scheme is devised particularly for the case of $|\mathcal{M}|=7$.
\begin{algorithm}
\caption{Adaptive decimation-enhanced FAID algorithm}
\begin{itemize}
\item[1)] Set $j=1$. Note that $\Phi_c$ will always be used to update messages at the check node. 
\item[2)] Initialize $\gamma_i=0$ $\forall v_i\in V$. 
\item[3)] Start the decimation procedure by passing messages for three iterations using $\Phi_v^{d}$. If the decoder converges within those three iterations, STOP.  
\item[4)] Apply decimation rule $\beta^{(1)}$ for every  $v_i\in V$. Then reset all messages to zero and set $q=0$.
\item[5)] Pass messages for two iterations using $\Phi_v^{d}$ for update at the non-decimated nodes. If the decoder converges within those two iterations, STOP. 
\item[6)] Apply decimation rule $\beta^{(2)[j]}$ only on nodes $v_i$ for which $\gamma_i=0$. Then reset all messages to zero. If $N_b>q$, $q=N_b$ and go  back to step 5, else go to step 7. 
\item[7)] Pass messages using $\Phi_v^{r}$ on the nodes $v_i$ for which $\gamma_i=0$.    
\item[8)] If decoder converges or has reached maximum allowed iterations, STOP. Else $j=j+1$. 
\item[9)] If $j>N_{\beta}$ STOP. Else go to step 2.
\end{itemize}
\end{algorithm}

{\it Remarks:} 1) The only criterion used by the decoder to decide when to use a more aggressive rule $\beta^{(2)[j]}$ on a given error pattern is whether the decoding has failed. 2) The reason for applying $\beta^{(1)}$ at the end of third iteration is that at least three iterations are required for a $\pm L_3$ to be passed. 3)The reason for the choice of every 2 iterations for applying $\beta^{(2)[j]}$ is because 2 iterations is small enough to help prevent the growth of wrong message strengths but sufficient to allow all levels in $\mathcal{M}$ to be passed. 

\subsection{Choice of $\Phi_v^{r}$ and $\Phi_v^{d}$}
For the proposed decoders, the map $\Phi_v^{r}$ is simply chosen to be the $\Phi_v$ of a particular FAID already known to be good on a given code, and for which we want to improve the guaranteed error correction capability. For the numerical results, $\Phi_v^{r}$ is chosen to be the $\Phi_v$ of a 7-level FAID defined by Table \ref{LUT244325}.

The choice of $\Phi_v^{d}$ on the other hand is non-trivial. It is designed based on analyzing messages that are passed within dense subgraphs that could potentially be trapping sets for a given FAID when errors are introduced in them under the isolation assumption. The rule is chosen under the premise that the growth of message strengths within the subgraph should be slow since many correct nodes in the subgraph would most likely be connected to error nodes, and multiple error nodes may be interconnected to each other in the subgraph (if the number of errors introduced is comparable to the size of the subgraph). Explicit design methods for $\Phi_v^{d}$ are not discussed in this paper, but we provide a particular $\Phi_v^{d}$ that was designed based on the above philosophy and used for the numerical results. It is an LT rule (see Section \ref{sect_Prelim}), so it can be described by assigning values to elements in $\mathcal{M}$, $\mathscr{T}$, and $\mathcal{Y}$. The map is defined with the following assignments; $L_1=1.1$, $L_2=2.3$, $L_3=6.6$, $T_1=0.8$, $T_2=2.8$, $T_3=4$, $\mrm{C}=1.5$. This was found to be a good rule for decimation.

\subsection{Analysis}
Due to page constraints, no proofs are provided (but they will be provided in the journal version of this paper). For the analysis, we assume that the all-zero codeword is transmitted which is valid since the decoders considered are symmetric. 
\begin{prop}
A node $v_i$ can receive a $\pm L_3$ from its neighbor $c_j$ in the first or second iteration after resetting the messages, only if all nodes in
$\mathcal{N}(c_j)\backslash v_i$ have been decimated. 
\end{prop}
\begin{lem}
If $\beta^{(2)[j]}(\mrm{C},L_3,-L_2,-L_2)=1$ and if $\forall c_k \in\mathcal{N}(v_i)$ all error nodes in $\mathcal{N}(c_k)\backslash v_i$ are non-decimated, then a correct node $v_i$ will be decimated if it receives an $L_3$ during a decimation round. 
\end{lem}

{\it Remark:} Note that $\beta^{(2)[j]}$ will always be defined so that $\beta^{(2)[j]}(\mrm{C},L_3,-L_2,-L_2)=1$ for any $j$ as explained in the next subsection. Also note how resetting messages at the end of each decimation round can help with decimating more correct nodes due to the above lemma. 
\begin{te}
If $\beta^{(2)[j]}(\mrm{C},L_3,-L_2,-L_2)=1$ and no error node is decimated, then any correct node in the residual graph $G'$ is connected to check nodes that have at least degree-two. 
\end{te}
\begin{cor}
If Theorem 1 holds and no error node in the residual graph $G'$ is connected to a degree-one check node, then $G'$ is a {\it stopping set}. 
\end{cor}

{\it Remark:} Note that if an error node in the residual graph $G'$ is connected to a degree-one check node, it would receive $L_3$ in every iteration for the remainder of the decoding (again assuming no error nodes are decimated), and this will most likely lead to a decoder convergence. Therefore, if no error node is decimated, the decoder is more likely to fail when the residual graph $G'$ is a stopping set (refer to \cite{richardsontrap} for details). 

The above remark is an important observation since we can now design the rules $\beta^{(1)}$ and the sequence $\{\beta^{(2)[j]}\}_j$ based on analyzing error patterns whose errors are entirely contained in the minimal stopping sets of a given code. For instance, if our goal is to correct up to $t$-errors, then we consider all error patterns up to a weight $t$ in the stopping sets in order to design $\beta^{(1)}$ and $\{\beta^{(2)[j]}\}_j$. 

If FAID $\mathscr{F}$ with $\Phi_v=\Phi_v^{r}$ has a critical number of $t$+$1$ on a stopping set whose induced subgraph is $H$, then $\mathscr{F}^{A}$ is  guaranteed to correct up to $t$ errors introduced in $H$ on the code if the residual graph is $H$. In other words, on a particular code, $\Phi_v^{r}$ is more likely to correct all error patterns up to weight-$t$ whose support lies in a stopping set in the code, if it has a critical number of $t+1$ on the stopping set.


\subsection{Discussion on design of decimation rules $\beta^{(1)}$ and $\beta^{(2)}$}
The design of $\beta^{(1)}$ involves selecting the triples that should be included in $\Xi^{(1)}$, which depends on the number of errors we are trying to correct and the type of harmful subgraphs present in $G$. $\beta^{(1)}$ should be chosen to be conservative enough so that no error nodes are decimated. On the other hand, the design of $\beta^{(2)}$ not only involves selecting the triples that should be included in $\Xi^{(2)}$, but also determining a specific ordering on the triples that will be included in subsets $\Xi^{(2)[j]}$ which determine the sequence of rules $\{\beta^{(2)[j]}\}_j$ used starting from the least conservative rule, and this is dependent on the structure of the code. Both rules can be designed by analyzing them on errors introduced in stopping sets of the code.

In order to specify the set $\Xi^{(1)}$, we just specify the message triples with the weakest values. For specifying $\Xi^{(2)}$ in a concise way, we shall introduce some notations. Let $\Xi^{(2)}$ be divided into two disjoint subsets, i.e., $\Xi^{(2)}=\Lambda \cup \Gamma$, where $\Lambda$ is a subset that contains all triples $(L_3,m_2,m_3)\in \mathcal{M}^3$ such that $m_2,m_3\geq -L_2$. Based on the analysis described previously, any $\Xi^{(2)[j]}$ defined should always have $\Lambda$ as its subset, regardless of the code. The subset $\Gamma$, which is dependent on the code, is an ordered set  whose ordering determines the subsets used to specify the sequence of rules $\{\beta^{(2)[j]}\}_j$.

\begin{table}[tp]
\begin{center}
\caption{Subset $\Gamma$ of $\Xi^{(2)}$ designed for $(732,551)$ code}

\begin{tabular}{ccc}
\begin{tabular}{|c|c|c|}
\hline $m_1$ & $m_2$ & $m_3$\\
\hline	$L_2$	&	$L_2$	&	$L_2$	\\
\hline	$L_2$	&	$L_2$	&	$L_1$	\\
\hline	$L_2$	&	$L_1$	&	$L_1$	\\
\hline	$L_2$	&	$L_2$	&	0	\\
\hline	$L_2$	&	$L_2$	&	$-L_1$\\
\hline					
\end{tabular} &
\begin{tabular}{|c|c|c|}
\hline $m_1$ & $m_2$ & $m_3$\\
\hline	$L_2$	&	$L_1$	&	0	\\
\hline	$L_1$	&	$L_1$	&	$L_1$	\\
\hline	$L_2$	&	0	&	0	\\
\hline	$L_2$	&	$L_1$	&	$-L_1$	\\
\hline					
\end{tabular} &
\begin{tabular}{|c|c|c|}
\hline $m_1$ & $m_2$ & $m_3$\\
\hline	$L_1$	&	$L_1$	&	0	\\
\hline	$L_2$	&	$L_2$	&	$-L_2$	\\
\hline	$L_2$	&	0	&	$-L_1$	\\
\hline	$L_1$	&	$L_1$	&	$-L_1$	\\
\hline	$L_1$	&	0	&	0	\\
\hline					
\end{tabular}
\end{tabular}
\label{decimatetable}\\
\caption{Subset $\Gamma$ of $\Xi^{(2)}$ designed for $(155,64)$ Tanner code}
\begin{tabular}{ccc}
\begin{tabular}{|c|c|c|}
\hline $m_1$ & $m_2$ & $m_3$\\
\hline	$L_2$	&	$L_2$	&	$L_2$	\\
\hline	$L_2$	&	$L_2$	&	$L_1$	\\
\hline	$L_2$	&	$L_2$	&	0	\\
\hline					
\end{tabular} &
\begin{tabular}{|c|c|c|}
\hline $m_1$ & $m_2$ & $m_3$\\
\hline	$L_2$	&	$L_1$	&	$L_1$	\\
\hline	$L_2$	&	$L_1$	&	0	\\
\hline					
\end{tabular} &
\begin{tabular}{|c|c|c|}
\hline $m_1$ & $m_2$ & $m_3$\\
\hline	$L_2$	&	$L_2$	&	$-L_1$	\\
\hline	$L_2$	&	$L_1$	&	$-L_1$	\\
\hline	$L_2$	&	0	&	0	\\
\hline					
\end{tabular}
\vspace{-2mm}
\end{tabular}
\label{decimatetable2}
\end{center}
\end{table}

\section{Numerical Results and Discussion}
\begin{figure}[tp]
\begin{center}
\includegraphics[angle=0, width=3.1in]{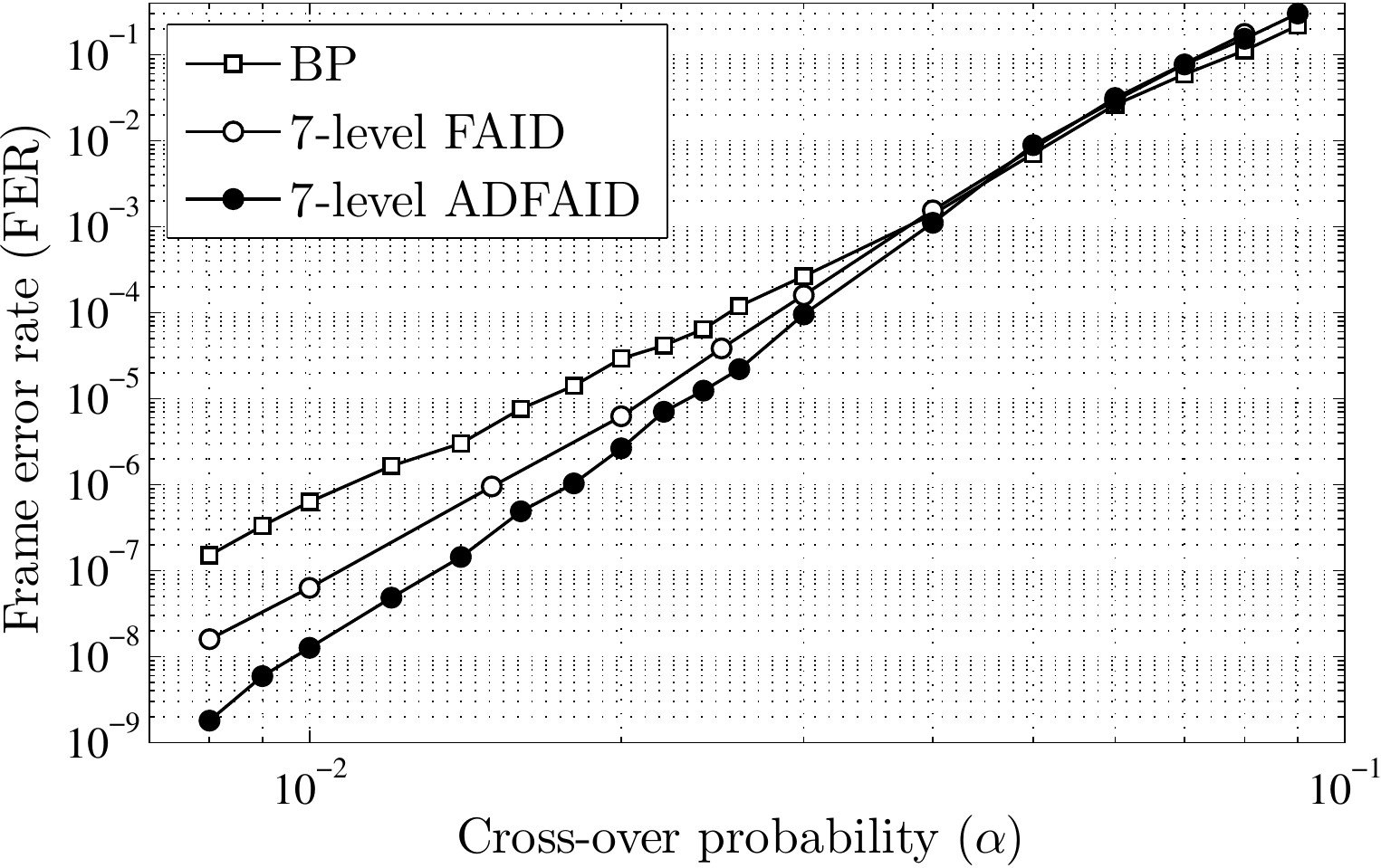}
\caption{FER performance comparison on the $(155,64)$ Tanner code}
\label{tannerplot}
\end{center}
\end{figure}

\begin{figure}[tp]
\begin{center}
\includegraphics[angle=0, width=3.1in]{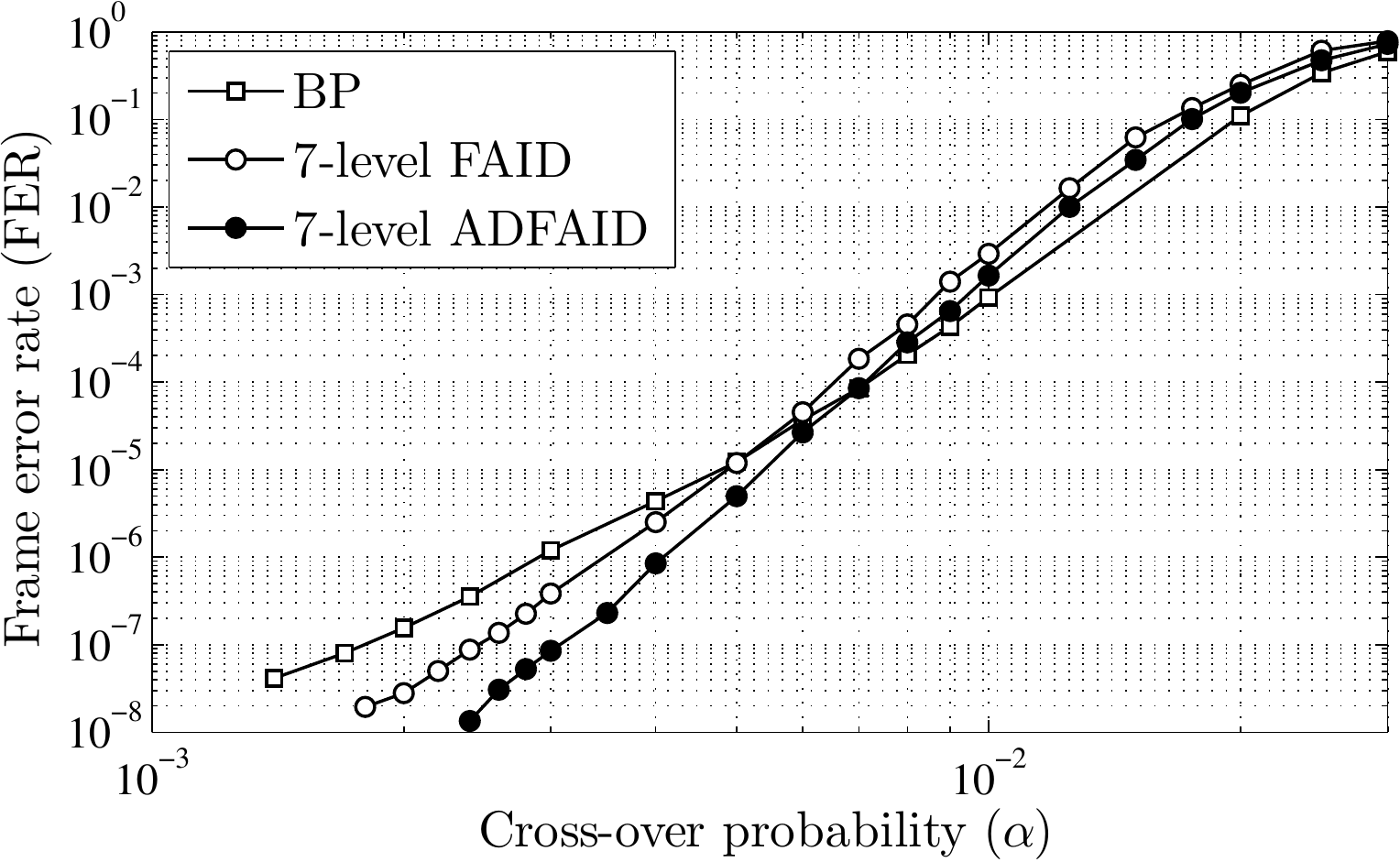}
\caption{FER comparison on the $(732,551)$ structured code with $d_{min}=12$}
\label{quasi732plot}
\end{center}
\vspace{-0.25in}
\end{figure}

Numerical results are provided in Fig. \ref{tannerplot} and Fig. \ref{quasi732plot} for two codes: the well-known $(155,64)$ Tanner code and a structured rate 0.753 $(732,551)$ code constructed based on latin squares \cite{DzungITW} with $d_{min}=12$. For the Tanner code, the set $\Xi^{(1)}$ contains all triples $(m_1,m_2,m_3)\in \mathcal{M}^3$ such that $(m_1,m_2,m_3)\geq (L_3,0,0)$ and $(m_1,m_2,m_3)\geq (L_2,L_2,L_1)$ (comparison is componentwise). For the high-rate structured code, $\Xi^{(1)}$ contains all triples such that  $(m_1,m_2,m_3)\geq (L_3,L_1,-L_3)$, $(m_1,m_2,m_3)\geq (L_3,-L_1,-L_1)$, and $(m_1,m_2,m_3)\geq (L_2,L_1,L_1)$. $|\Xi^{(1)}|=12$ for the Tanner code and $|\Xi^{(1)}|=24$ for the $(732,551)$ code. The $\Gamma$ sets in $\Xi^{(2)}=\Lambda \cup \Gamma$ for the structured code and the Tanner code are shown in Tables \ref{decimatetable} and \ref{decimatetable2} respectively. The cardinalities of the subsets of $\Xi^{(2)}$ used by each of the two codes are 
$\{|\Xi^{(2)[j]}|\}_j=\{24,25,26,27,28,29,30,31,32,33,34,35\}$ and $\{|\Xi^{(2)[j]}|\}_j=\{23,25,26,27,29\}$ respectively. The maximum number of iterations allowed for BP and 7-level FAID, and for $\Phi_v^{r}$ of the 7-level ADFAID, was 100.

The significant improvement in the slope of the error floor by using the 7-level ADFAID is evident. For the Tanner code, it was verified that all 6-error patterns are corrected by the 7-level ADFAID while the 7-level FAID corrects all 5-errors and BP fails on 5-errors. For the high-rate structured code, no failed 5-error patterns were found in the region of simulation shown in Fig. \ref{quasi732plot}, which is significant since the code has $d_{min}=12$. This shows that for certain high-rate codes whose graphs are relatively dense and for which it becomes difficult to ensure high $d_{min}$ in the code, the FAIDs with adaptive decimation can possibly come close to achieving the guaranteed error correction of maximum likelihood decoding. Note that the 7-level ADFAIDs are still 3-bit message passing decoders which have reasonable complexity, and that is still lower than BP.

%

%
%
\section*{Acknowledgment}
This work was funded by NSF grants CCF-0830245 and CCF-0963726, and by Institut Universitaire de France grant.

\end{document}